\documentclass[sn-apa,authoryea]{sn-jnl}

\usepackage{graphicx}%
\usepackage{multirow}%
\usepackage{amsmath,amssymb,amsfonts}%
\usepackage{amsthm}%
\usepackage{mathrsfs}%
\usepackage[title]{appendix}%
\usepackage{xcolor}%
\usepackage{textcomp}%
\usepackage{manyfoot}%
\usepackage{booktabs}%
\usepackage{algorithm}%
\usepackage{algorithmicx}%
\usepackage{algpseudocode}%
\usepackage{listings}%
\usepackage{threeparttable}
\usepackage{rotating}
\usepackage[section]{placeins}
\usepackage{anyfontsize}
\usepackage{comment}
\usepackage[english]{babel}
\usepackage{caption}
\usepackage{subcaption}
\usepackage{hyperref} 
\hypersetup{
    colorlinks=true,                          
    linkcolor=blue, 
    citecolor=red, 
    urlcolor=blue  }
\usepackage{footnote}
\makesavenoteenv{table} 
\newtheorem{rem}{Remark}

\definecolor{GreenForest}{RGB}{0, 140, 80}

\begin{document}

\title{Unveiling Crowdfunding Futures: Analyzing Campaign Outcomes through Distributed Models and Big Data Perspectives}

\author*[1]{\fnm{Giuseppe} \sur{Pipitò}}\email{g.pipito@studenti.uniba.it}

\author[2]{\fnm{Emanuele} \sur{Macca}}\email{emanuele.macca@unict.it}

\affil[1]{\orgdiv{Department of Computer Science}, \orgname{University of Bari}, \orgaddress{\city{Bari}, \country{Italy}}}

\affil[2]{\orgdiv{Department of Mathematics and Computer Science }, \orgname{University of Catania}, \orgaddress{\city{Catania}, \country{Italy}}}


\abstract{
Crowdfunding has emerged as a widespread strategy for startups seeking financing, particularly through reward-based methods. However, understanding its economic impact at both micro and macro levels requires thorough analysis, often involving  advanced studies on past campaigns to extract insights that aiding companies in optimizing their crowdfunding project types and launch methodologies. Such analyses are often beyond the scope of basic data analysis techniques and frequently demand advanced machine learning tools, such as distributed computing, due to the large volume of data involved. This study aims to investigate and analyse the targets of reward-based crowdfunding campaigns through machine learning techniques, employing distributed models and structures. By harnessing the power of distributed computing, it unravels intricate patterns and trends within crowdfunding data, thereby empowering companies to refine their strategies and enhance the efficacy of their funding endeavors. Through this multifaceted approach, a deeper understanding of the economic dynamics underlying crowdfunding ecosystems can be attained, fostering informed decision-making and sustainable growth within the startup landscape.

\vspace{0.5cm}
\noindent\textbf{Plain English Summary: } "Unraveling Crowdfunding Secrets: Distributed Analytics Reveals Surprising Insights For Startup Strategies. \#Crowdfunding \#DistributedAnalytics \#MachineLearning"

This study employs advanced machine learning techniques, including distributed computing, to dissect the targets of reward-based crowdfunding campaigns. By analyzing extensive datasets from past campaigns, we uncover nuanced patterns and trends, enabling startups to optimize their funding projects and launch methods. Surprisingly, our research reveals previously unseen correlations and predictive indicators, revolutionizing how companies approach fundraising. This breakthrough empowers startups to make informed decisions, enhancing their chances of success in the competitive landscape of crowdfunding. Implications for society: Startups can leverage these insights to refine their crowdfunding strategies, maximizing their fundraising potential, accelerating their growth trajectory and contributing to develop creative manifestations of the human being (visual art, game, book, cinema and theatre).
}

\keywords{Reward-based crowdfunding, Machine learning, Predictive modeling, Big Data, Distributed model}



\maketitle

\section{Introduction}
Crowdfunding has been experiencing a substantial surge in recent years, providing a novel means of financial support for a wide range of projects [and ventures] across various sectors as an alternative to traditional investments. In this regard, a comprehensive exploration of reward crowdfunding campaigns, employing advanced Big Data analytics and distributed computing architectures to train machine learning models is useful. The study encompasses a plentiful dataset scraped from \href{https://www.kickstarter.com/}{Kickstarter}, a crowdfunding platform. Usually, the crowdfunding campaigns could be distinguished into four primary categories: \textit{equity, lending, reward}, and \textit{donation}. Such categories exhibit distinctive characteristics and understanding these nuances is paramount for an analytical journey. In particular, \textit{equity} and \textit{lending} involve financial instruments offering the possibility of yielding profitable investments~\cite{Belleflamme2014,Colombo2015}; \textit{donation} is characterized by backers contributing funds without any expectation of material returns~\cite{Meer2013}; and \textit{reward}, which is the focus of the study, allows backers to pledge funds to a campaign in exchange for non-financial rewards or products. This category encompasses a broad spectrum of projects, ranging from technological innovations to creative works~\cite{zhu2022Proximal}. Effectively, the paper aims the outcome prediction of a crowdfunding campaign reward, as \cite{Etter2013}, adopting techniques implemented on advanced distributed framework.

For efficient analysis, the reward crowdfunding paradigm, being the most extensively studied \cite{Mol2014,Kraus2016} and widely used among the four categories \cite{Belleflamme2015}, requires distributed computing architectures \cite{Xing2016} due to the sheer volume of data generated by campaigns on platforms, as Kickstarter \cite{Wan2022}. Nevertheless, the actually state-of-art has predominantly focused on traditional undistributed models such as linear and logistic regression, and random forests for predicting campaign outcomes based on less campaign records, see \cite{Koc2016,Song2019,Wan2020} respectively.
By the way, the article uses an innovative perspective in this field.

In contrast to many studies that employ binary target variables to predict crowdfunding success, e.g.~\cite{Zho2022,Shneor2020}, this work accentuates the use of non-binary categorical target variables. The considered dataset encompasses campaigns classified into five distinct outcome categories: “Success”, “Failed”, “Canceled”, “Suspended”, and “Live”. To enhance the analysis, we initially consider the categories and subsequently group them into two overarching classes: “Successful” and “Not Successful”.

The article aims to provide a detailed insight into the methodologies, data preprocessing techniques, and machine learning algorithms employed to predict the outcomes of reward crowdfunding campaigns, with a particular emphasis on the multi-class prediction aspect. By leveraging distributed computing and conducting extensive data analysis, our research endeavors to contribute to the evolving field of crowdfunding analytics, offering a more comprehensive understanding of the factors influencing campaign success [and failure].

It is evident that crowdfunding is a dynamic landscape, in which reward crowdfunding playes a pivotal role. The decision to explore non-binary categorical target variables in campaign outcome prediction reflects a commitment to deepening the analysis within this domain. The application of distributed computing and advanced machine learning techniques presents a promising avenue for addressing the complexities and scale of data associated with reward crowdfunding campaigns on platforms, as the case of Kickstarter.

The rest of this paper is organized as follows: 
Section~\ref{sec:materials} presents the dataset focusing on reward crowdfunding campaigns with a preliminary analysis. In Section~\ref{sec:methods},the methodology, such as modeling and preprocessing, and the distributed models adopted for the study are described. Numerical results, beckon accuracy and F1-score, are reported in Section~\ref{sec:numerical_results}, showing the performances of the adopted models.
Conclusions, perspectives and potential implications are drawn in Section~\ref{sec:conclusions}.

\section{Materials}\label{sec:materials}
The dataset under analysis has been released on \hyperlink{https://www.kaggle.com/datasets/kemical/kickstarter-projects}{Kaggle} by Mickaël Mouillé\footnote{The dataset is released under the CC BY-NC-SA 4.0  (Creative Commons Attribution - Non-Commercial - Share-Alike) license, which permits its distribution, modification, and creation of derivative works, as long as appropriate credit is given and the same license is applied.}, a crowdfunding enthusiast who has been familiar with the Kickstarter platform since its public release. It's important to note that the dataset is not officially released from Kickstarter but is obtained through scraping techniques. While it may not include the latest campaigns, its availability offers an excellent opportunity for analysis and exploration. Nevertheless, there have been previous studies on Kickstarter campaign data \cite{blan2022Extracting,Raf2023}.

It possesses immense potential for public administrations and research centers to acquire valuable insights into social phenomena. It can empower informed decision-making concerning welfare policies and cultural initiatives, guiding funding decisions and facilitating the adoption of data-driven approaches.

\subsection{Business Understanding}
As widely known, crowdfunding is a process that enables a varying number of individuals to finance and support projects, businesses, or initiatives through an online platform. There are several types of crowdfunding, including:
\begin{itemize}
    \item \textit{Reward} crowdfunding, the most common  crowdfunding form in terms of projects number and supporters in which backers receive rewards for their contribution. The rewards can be product copy or service being developed, special acknowledgements, unique experiences, or other incentives.

    \item \textit{Equity} crowdfunding, in which investors receive a share of ownership in the company in exchange for their contribution. It is commonly used by start-ups or businesses seeking investments to expand their operations. 
    
    \item \textit{Lending} crowdfunding, in which investors  provide loans -- often with high-interest rates -- to companies or initiatives. The crowdfunding platform acts as an intermediary between the lender and the borrower. It is prevalent in real estate ventures. 

    \item  \textit{Donation} crowdfunding, in which backers raise for charity.     
\end{itemize}

Out of the various crowdfunding types mentioned, this study lies in reward-based crowdfunding. Accordingly, the Kickstarter platform has been considered. Such platform, founded in 2009, is a prominent reward-based crowdfunding website that has left a substantial imprint on society. 

The Kickstarter campaigns can be categorized into five temporary or definitive states: Live, Successful, Failed, Canceled, and Suspended.
\begin{itemize}
    \item Live: This temporary state is assigned immediately after the campaign is launched\footnote{ Some campaigns considered in the dataset might not have concluded at the time of data acquisition.}. Once the campaign ends, this temporary state transitions into one of the four definitive states.

    \item Successful: Projects that have successfully achieved their predetermined funding goal within the specified timeframe are categorized as successful.

    \item Failed: Campaigns that have not attained their predetermined funding goal within the specified timeframe are considered failed.

    \item Canceled: Projects that have been permanently canceled by the creators\footnote{A canceled project can potentially be relaunched, but based on the available data, it is challenging to determine whether a project has previously been canceled or not.}.

    \item Suspended: Projects that have been permanently suspended by the platform due to violations of the Kickstarter's policies.
\end{itemize}

\subsection{Data Mining Goals}
In the realm of analyzing Kickstarter campaigns, the data mining objective is State classification. Indeed, the primary aim is to classify the different states of Kickstarter campaigns, enabling the derivation of significant insights into various characteristics of the phenomenon. By categorizing campaigns into distinct states, we can gain valuable information that illuminates important aspects of the crowdfunding landscape. These insights can be used in decision-making processes, providing a data-driven foundation for strategic choices.

By pursuing this goal, our data mining endeavors aim to extract meaningful patterns, uncover valuable insights, and provide predictive capabilities from the Kickstarter dataset. From a business point of view, this empowers stakeholders to make informed decisions, optimize their strategies, and understand the key factors that contribute to campaign success.

\subsection{Data Understanding}
During the preliminary analysis, several challenges have been encountered.

Firstly, it has been identified a remarkable challenge related to the presence of two distinct features, namely usd\_pledged and usd\_pledged\_real, both of which aim to capture the same characteristic of the crowdfunding phenomenon. This redundancy introduces a potential source of confusion in the interpretation of the dataset, analyzed in Sec.~\ref{sec:methods}.

Furthermore, inherent complexities are associated with the crowdfunding phenomenon itself. Notably, the values of currency, a crucial aspect in our dataset, exhibit significant fluctuations throughout the duration of campaigns. Consequently, retrospective conversions are subject to approximation, and the precision of these conversions varies. Nevertheless, this particular aspect has not been engaged in this article due to the extreme issue complexity. 

These challenges highlight the complexities involved in understanding and interpreting the dataset. While efforts have been made to preprocess and integrate the data effectively, certain uncertainties and limitations persist due to the nature of the crowdfunding phenomenon and currency fluctuations.

\subsubsection{Variables}
For sake of clarity, a list of variables has been provided in Table~\ref{tab:variables} with a short description for a better understanding of both, the analysed phenomenon and the considered dataset.

\begin{rem}
\label{rem_variables}
Some noteworthy observations arise concerning the nature of the variables within the dataset. These observations are crucial for understanding the limitations and scope of our analysis. 

Firstly, the dataset lacks variables that explicitly indicate the promised reward or furnish information about its specific characteristics. Consequently, the absence of such information restricts the ability predicting campaign outcomes or conducting an independent analysis focused on the nature of the rewards promised.

Moreover, do not exist variables that could signify whether the expected rewards have been delivered to the campaign backers or provide insights into the distribution process itself. 

These observations underscore certain limitations in the dataset, specifically regarding the absence of key variables that would enhance the depth and breadth of the analysis in predicting outcomes and understanding the reward fulfillment dynamics within crowdfunding campaigns.
\end{rem}

\begin{table}[t]
\centering
\begin{tabular}{lll}
\toprule
Variable Name      & Variable Meaning                                   & Type         \\ 
\midrule
name               & Name of the Kickstarter project                     & String               \\ 
main\_category      & Main category of the crowdfunding campaign          & String               \\ 
category           & Subcategory of the crowdfunding campaign            & String               \\ 
launched           & Date and time of campaign launch                    & Timestamp \\ 
deadline           & Date of campaign deadline                           & Date    \\ 
state              & State of the campaign    & String               \\ 
backers            & Number of campaign backers                          & Int                  \\ 
currency           & Currency used for funding                           & String               \\
country            & Country                                             & String               \\
goal               & Campaign goal in USD                                & Float $\rightarrow$ Int         \\ 
usd\_goal\_real      & Campaign goal converted to USD (via \hyperlink{https://fixer.io/}{Fixer.io API}) & Float           \\ 
pledged            & Amount pledged in USD (refunded in case of failure) & Float \\ 
usd\_pledged        & Pledged amount converted to USD by Kickstarter       & Float                \\ 
usd\_pledged\_real   & Pledged amount converted to USD (via \hyperlink{https://fixer.io/}{Fixer.io API}) & Float          \\ 
\bottomrule
\end{tabular}
\vspace{0.1cm}
\caption{List of dataset variables, a short description including them types.}
\label{tab:variables}
\end{table}
\vspace{0.5cm}
\begin{rem} \textbf{Known relationships between variables:}
The hierarchical relationship between main category and subcategory variables is crucial in comprehending the organizational structure of crowdfunding campaigns, wherein the former delineates the overarching thematic group, and the latter provides nuanced information about the project's specific domain within that main category. The backers variable, quantifying the number of individuals financially supporting a campaign, assumes significance as an indicator of audience engagement, with a requisite minimum number of backers often deemed essential for campaign success. Financial aspects are encapsulated by variables such as goal, usd\_goal\_real, pledged, usd\_pledged, and usd\_pledged\_real, collectively representing the campaign's funding goal and the amount of money pledged. Noteworthy is the equivalence of pledged, usd\_pledged, and usd\_pledged\_real when the currency is USD. Additionally, specific variables, such as name, offer an avenue for exploring the potential influence of nomenclature characteristics on campaign outcomes.
\end{rem}

\subsection{Preliminary analysis}
The primary target variable in the dataset is State, which denotes the potential states of a campaign. The count and percentage of campaigns grouped by State are shown in Table~\ref{tab:state}.
\begin{table}[ht]
\centering
\begin{tabular}{lrr}
\toprule
State      & Count         &    Percentage      \\ 
\midrule
Failed & 196,945 & 52.72\% \\
Successful & 133,324 & 35.69\% \\
Canceled & 38,665 & 10.35\% \\
Live & 2,795 & 0.75\% \\
Suspended & 1,841 & 0.49\% \\
\bottomrule
\end{tabular}
\caption{Distribution of campaigns grouped by state.}
\label{tab:state}
\end{table}
\begin{figure}[!ht]
     \centering
     \includegraphics[width=0.8\textwidth]{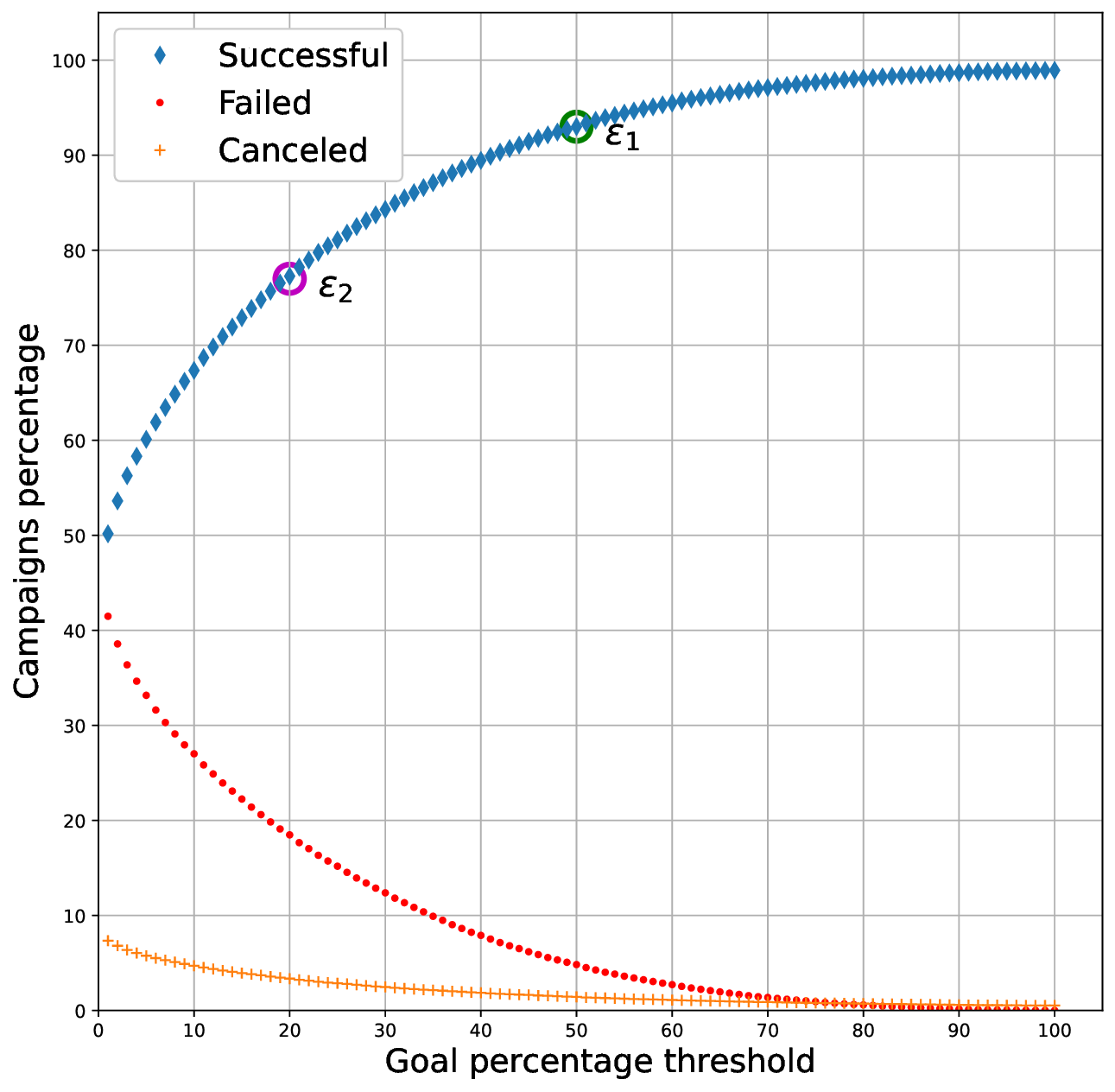}          
     \caption{Relations between Successful, Failed and Canceled. The points $\varepsilon_1$ and $\varepsilon_2$ represent the percentage of successful campaigns that have collected at least $50\%$ and $20\%$ of the goal, respectively.}
     \label{Fig:threshold_1}
\end{figure}
\begin{figure}[!ht]
     \centering
     \begin{subfigure}[b]{1.01\textwidth}
         \centering         \hspace{-0.5cm}\includegraphics[width = \textwidth]{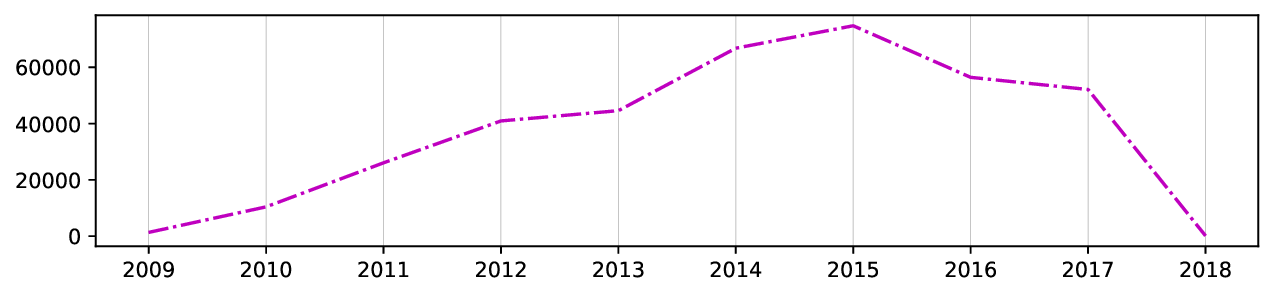}
         \caption{New projects per year.}
         \label{New_Projects}
     \end{subfigure}     
     \begin{subfigure}[b]{0.975\textwidth}
         \centering         \hspace{0.25cm}\includegraphics[width=\textwidth]{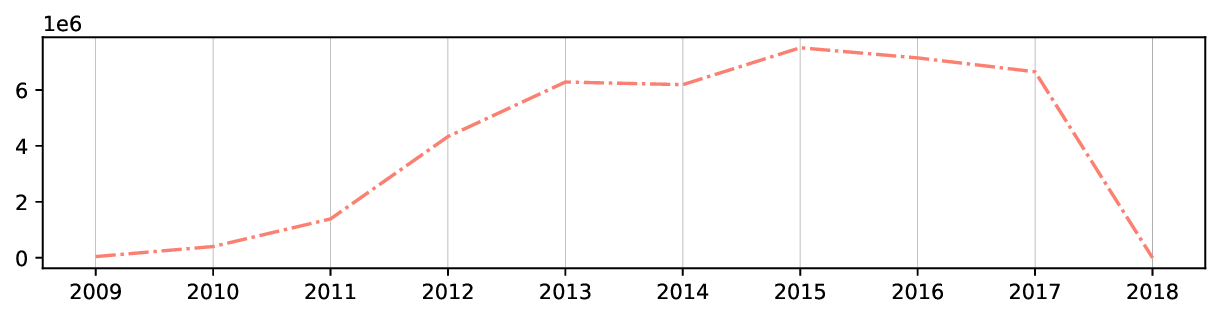}
         \caption{Backers per year in million.}
         \label{Backers_M}
     \end{subfigure}     
     \begin{subfigure}[b]{0.975\textwidth}
         \centering
        \hspace{0.25cm} \includegraphics[width=\textwidth]{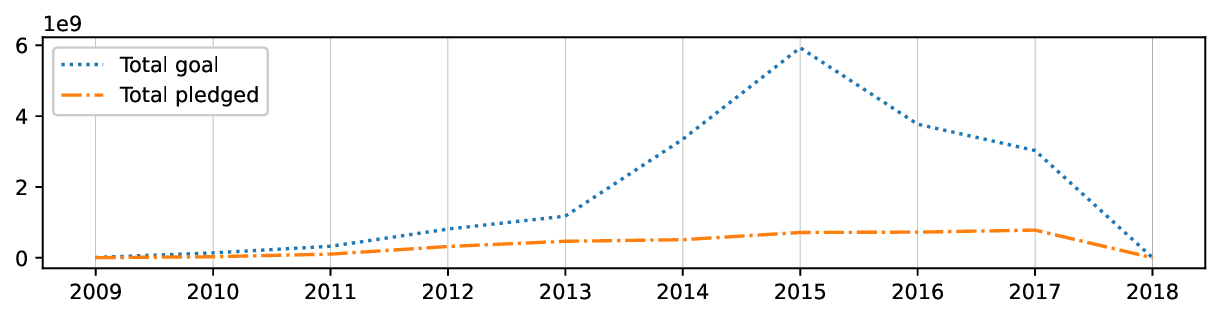}
         \caption{Total money goal required and total pledged promised per year in billion.}
         \label{Goal_vs_pledged_B}
     \end{subfigure}  
     \begin{subfigure}[b]{0.98\textwidth}
         \centering
         \includegraphics[width=\textwidth]{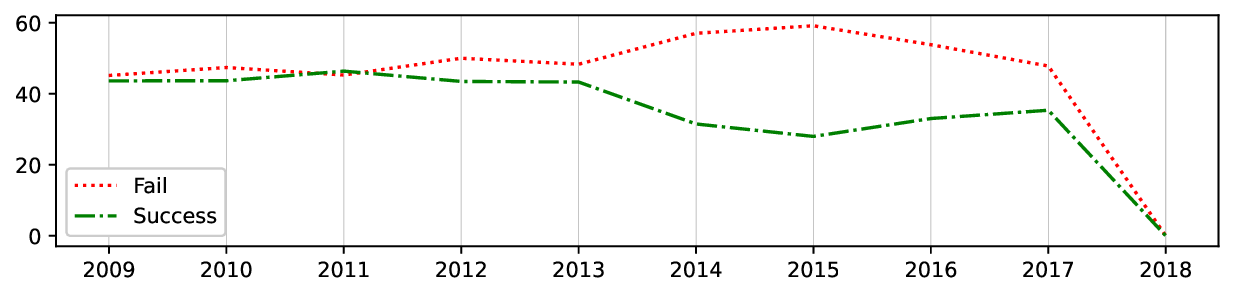}
         \caption{Fail and Success rates per year.}
         \label{Fail_vs_Success_rate}
     \end{subfigure}  
     \caption{Totals per year: new open projects (a); numbers of backers expressed in million (b); amount of money required and promised in billion (c); and, percentage of projects failed and successful (d).}
     \label{Fig:Totals_per_year}
\end{figure}
In particular, Failed campaigns constitute the largest portion, accounting for $52.72\%$ of the dataset; Successful campaigns follow, making up $35.69\%$ of the dataset; Canceled campaigns represent $10.35\%$ of the dataset; Live campaigns have a very small presence, comprising only $0.75\%$ of the dataset; and Suspended campaigns have the smallest representation, with just $0.49\%$ of the dataset.

Temporarily, the states Failed, Successful, and Canceled have been considered, and they have been compared with the percentage of the prearranged goal threshold.

Figure~\ref{Fig:threshold_1} displays that, once approximately $50\%$ of the amount is surpassed, about $93\%$ of the campaigns will achieve the established goal, marked as $\varepsilon_1$, and approximately $5\%$ of the campaigns will not reach the goal, while $2\%$ will be definitively canceled. Similar considerations can be made when surpassing the $20\%$ of the amount, marked as $\varepsilon_2$, with lower percentages for Successful campaigns and higher percentages for Failed and Canceled campaigns.
\vspace{0.25cm}
\begin{rem}
    It is worth noting that considering only the three states, the sum of the percentages almost-always equals $100\%$, since there exist the other two minority states. 
\end{rem}
\vspace{0.25cm}
\begin{rem}    
From a careful analysis of Figure~\ref{Fig:threshold_1}, it's evident that as soon as the campaign starts receiving funding ${\rm (goal\;\; percentage}>0)$, the success rate is 50\%. Nevertheless, and crucially, since there is no information between the percentage of the preset goal threshold and the number of days remaining until the campaign deadline, these observations hold only qualitative statistical significance.
\end{rem}
\vspace{0.25cm}

Figure~\ref{Fig:Totals_per_year} represents the number of new projects opened \ref{New_Projects}, the number of backers in millions \ref{Backers_M}, the required and promised amounts\footnote{Promised amounts refer to campaigns classified as Successful.} in billions \ref{Goal_vs_pledged_B}, and the success percentage compared to the failure percentage per year \ref{Fail_vs_Success_rate}. 
In particular, there is a discernible growth in the number of projects until the year 2015, at which point a decline in project numbers becomes apparent in subsequent years. In contrast, the backers plot exhibits growth until 2013, after which it stabilizes, maintaining a relatively consistent pattern. While the pledged amounts exhibit a lack of significant fluctuations over the analyzed period, there is a conspicuous surge in requested amounts in the years 2014 and 2015. However, this surge is swiftly followed by a notable decrease in the ensuing years. Finally, examining the success and failure rates of projects reveals consistent patterns until 2013. Anyway, a deviation occurs in 2014 and 2015, marked by an increase in the failure rate and a simultaneous decrement in the success rate.

One plausible interpretation of these trends could be traced to the platform's growth. It appears that until 2013, both the number of projects and supporters were on an upward trajectory. However, in 2014, there was a substantial increase in project numbers while supporter numbers remained stationary. This suggests that marketing efforts directed at project creators were more successful than those aimed at attracting sponsors. Consequently, the increase in the failure rate could be attributed to the rise in requested amounts, despite the pledged amounts remaining relatively stable.  This trend was further accentuated in the ensuing year, 2015. Subsequently, a decline in the number of projects and absolute requested amounts led to a subsequent increase in the success rate.

Understanding these trends can be useful:
\begin{enumerate}
    \item For the platform or its competitors, as it becomes evident that engaging a larger number of backers is equally essential as enticing more creators to actively utilize the platform.
    
    \item For project creators, recognizing that the platform does not significantly aid in attracting new backers could emphasize the importance of independent and well-targeted marketing campaigns to inform potential investors.
\end{enumerate}

\section{Methods}\label{sec:methods}
Due to the large amount of data, the methods have been implemented through Apache Spark \cite{Zaharia2010,Zaharia2012,Zah2016}, a multi-language engine for executing data engineering, data science, and machine learning on single-node machines or clusters. 

\subsection{Preprocessing}
Due to inconsistent information, some adjustments were necessary to handle the data effectively. Below, the removed observations list and it explanations:
\begin{itemize}
    \item 5 observations with a Successful state did not obtain funding equal to or exceeding the preset goal.
    \item 6 observations with a Failed state received funding higher than the requested amount.
    \item 3 observations were eliminated as they had 0 backers but still received funding.
\end{itemize}
Moreover, it was noticed that the variables usd\_pledged and usd\_pledged\_real did not display the same values in USD currency. This behavior is highly illogical since, in the case of USD currency with no currency conversion, the values must be identical. As a consequence, the usd\_pledged variable was removed from our dataset and unconsidered in the analysis.
\vspace{0.25cm}
\begin{rem}
    One option would have been to discard the observations affected by this issue. Nevertheless, we discourage this approach for two reasons: first, it is only relevant to USD currency due to the lack of currency conversion; secondly, these observations comprise approximately 12\% of the entire dataset, a substantial portion.
\end{rem}

\paragraph{Features engineering}
In order to manage the date-type variables, we extracted year, day of week, hour, month and trimester from both launched and deadline variables; the duration has been computed by the difference between deadline and launched; continent has been generated by country as three string values: “Europe”, “America” and “Asia/Oceania”; pledged\_per\_backer and usd\_pledged\_real\_per\_backer\footnote{ usd\_pledged\_real\_per\_backer is the ratio between real pledged and backer in USD.} have been generated from backers, pledged and usd\_pledged\_real; from name its length, the number of words composing it, the number of its capital letters, the number of its alphanumerical symbols and the number of characters that are digits have been derived.

Furthermore, some boolean features have been generated depending on the presence of specific word in the name such as: new, first, world, help, project, canceled and suspended. In selecting individual words for feature inclusion, preference has been given to terms such as \textit{new} and \textit{first} (commonly associated with innovative concepts), \textit{world} (typically indicative of ideas with global implications), \textit{help} (linked to philanthropic notions), and \textit{project} (associated with well-organized ideas). It is essential to acknowledge potential semantic ambiguities that may not be easily identified and addressed, such as the occurrence of \textit{new} in the context of \textit{New York}, a frequent and notable bigram. This decision is inherently subjective, introducing the possibility of biases that should be taken into account. 

\paragraph{Data integration}
The data have been integrated with macroeconomics index such as Gross Domestic Product (GDP) and Human Development Index (HDI) to asses their impact on a microeconomic phenomenon as crowdfunding campaign, see \cite{Bolt2020,Ros2014}. This integration involves adding two new variables, related to index (GDP and HDI) for each respective year and country.  

\begin{table}[ht]
\begin{subtable}{0.49\textwidth}
\centering
\begin{tabular}{lrr}
\toprule
Country & Mean ($\$$) & Std ($\%$) \\
\midrule
AUS & 47463 & 3.6 \\
AUT & 41338 & 2.3 \\
BEL & 38318 & 2.1 \\
CAN & 42959 & 3.4 \\
CHE & 58879 & 2.9 \\
DEU & 43470 & 4.9 \\
DNK & 44132 & 2.8 \\
ESP & 31122 & 2.1 \\
FRA & 36840 & 2.3 \\
GBR & 36148 & 3.7 \\
HKG & 46363 & 6.5 \\
IRL & 53081 & 11.1 \\
ITA & 33889 & 2.3 \\
JPN & 36364 & 4.4 \\
LUX & 54911 & 3.3 \\
MEX & 15614 & 4.9 \\
NLD & 44808 & 3.1 \\
NOR & 81545 & 3.0 \\
NZL & 33208 & 4.1 \\
SGP & 63182 & 8.0 \\
SWE & 42991 & 4.1 \\
USA & 51526 & 4.3 \\
\bottomrule
\end{tabular}
\caption{Statistics for GDP per capita grouped by country for all the range year 2009-2018.}
\label{tab:dati_gdp}
\end{subtable}
\hspace{0.6cm}
\begin{subtable}{0.49\textwidth}
\centering
\begin{tabular}{lrr}
\toprule
Country & Mean ($\$$) & Std ($\%$) \\
\midrule
AUS & 0.9306 & 0.7 \\
AUT & 0.9083 & 0.7 \\
BEL & 0.9203 & 0.9 \\
CAN & 0.9222 & 0.9 \\
CHE & 0.9492 & 0.8 \\
DEU & 0.9352 & 0.8 \\
DNK & 0.9309 & 1.3 \\
ESP & 0.8821 & 1.5 \\
FRA & 0.8877 & 1.1 \\
GBR & 0.9191 & 1.0 \\
HKG & 0.9259 & 1.7 \\
IRL & 0.9151 & 1.6 \\
ITA & 0.8843 & 0.5 \\
JPN & 0.9106 & 1.2 \\
LUX & 0.9172 & 0.7 \\
MEX & 0.7619 & 1.5 \\
NOR & 0.9497 & 0.9 \\
NLD & 0.9285 & 0.9 \\
NZL & 0.9282 & 0.7 \\
SGP & 0.9227 & 1.5 \\
SWE & 0.9262 & 1.6 \\
USA & 0.9177 & 0.6 \\
\bottomrule
\end{tabular}
\caption{Statistics for HDI per capita grouped by country for all the range year 2009-2018.}
\label{tab:dati_hdi}
\end{subtable}
\caption{Statistics on state economic data grouped for the period 2009-2018.}
\label{tab:dati_economici}
\end{table}

\begin{rem}
The GDP and HDI data pertain to the year when the campaign initiates. Nevertheless, this represents an approximation and may not accurately reflect the comprehensive data for the entire campaign, particularly in cases where a campaign spans multiple consecutive years. Despite these approximations at a first-order level, they do not exert a significant influence on the classification. The error introduced at the outset due to such approximations is relatively minimal since the global standard deviations, in percentage, are small values, as evidenced by the statistics presented in Table \ref{tab:dati_gdp}-\ref{tab:dati_hdi}. The local two-years variations are omitted due to its excessive volume. In order to achieve a more accurate approximation of GDP and HDI data, it would be sufficient to weight the GDP or HDI values based on the actual days that the campaign is active in a given year. In mathematical terms, let $gg_i$ represent the number of days in which the campaign is active in year $i$, and $gg_{i+1}$ represent the number of days in which the campaign is active in year $i+1$, similarly, $gg_{i+1} = dd - gg_i$, where $dd$ denotes the total duration of the campaign. We can define, the new GDP value as 
\[ gdp_{\text{new}_j} = \frac{gg_i \cdot gdp_i + gg_{i+1} \cdot gdp_{i+1}}{dd} \]
where $j = j(i)$. Similarly the formulation for the new HDI value \[ hdi_{\text{new}_j} = \frac{gg_i \cdot hdi_i + gg_{i+1} \cdot hdi_{i+1}}{dd}. \]
These formulas do not contains campaigns distributed on 3 or more years.
\end{rem}

\paragraph{Data preparation for models}
Due to the label Live refers to campaign still ongoing, it was deemed appropriate to remove it from the analysis. For this reason, two new sets of classes have been adopted: 
\begin{align*}
\label{Group_name}
\mathcal{I}_1&:=\{\rm{Successful},\rm{Fail},\rm{Canceled},\rm{Suspended}\},  \\ \nonumber  \mathcal{I}_2&:=\{\rm{Successful},\rm{Not\ Successful}\},
\end{align*}
$\mathcal{I}_1$ with the original labels and $\mathcal{I}_2$ with old label Successful and new label Not Successful that incorporate Fail, Canceled and Suspended. This choice is made to represent the variable state in a Boolean sense, emphasizing the Successful class.

Since the label distribution is unbalanced (see Table~\ref{tab:state}), weights have been introduced to mitigate the impact of this phenomenon. These weights $w_{i,j}$ have been computing in a heuristic way, according to \cite{King2002}, as:
\begin{equation}
    \label{weight_distribution}
    w_{i,j} := \frac{\sum_{\ell=1}^{N_j}\lambda_\ell}{N_j \lambda_i}    
\end{equation}
where $i\in\mathcal{I}_j$ for $j=1,2;$ $\lambda_i$ are the frequencies of a class $i\in\mathcal{I}_j$; and, $N_j$ is the size of $\mathcal{I}_j$.

Observe that $$\sum_{i=1}^{N_j}\lambda_iw_{i,j} = \sum_{i=1}^{N_j} \frac{\sum_{\ell=1}^{N_j}\lambda_\ell}{N_j} = \frac{\sum_{\ell=1}^{N_j}\lambda_\ell}{N_j} \sum_{i=1}^{N_j} 1 = \sum_{\ell=1}^{N_j}\lambda_\ell=\sum_{i=1}^{N_j}\lambda_i.$$

In order to finally prepare the data for the analysis, they have been trained on the training sets $\mathcal{T}_1$ and $\mathcal{T}_2$, and valuated on the test sets $\mathcal{V}_1$ and $\mathcal{V}_2$ adopting the stratified sampling (see next paragraph for details).

\paragraph{Stratified sampling} \label{strati_sampl}
Stratified sampling involves dividing a population into homogeneous subgroups (strata), and then randomly sampling within each stratum. This approach ensures that each subgroups is represented leading to more accurate estimation.

In this case\footnote{In the undistributed scenario, there exist different implemented strategy for the stratified sampling. This is not the case for the distributed scenario that requires manual and not straightforward implementations.}, the subgroups are the distinct values of the target variable: $\mathcal{I}_1$ and $\mathcal{I}_2$ defined above. Consequently, the training sets, $\mathcal{T}_1$ and $\mathcal{T}_2$, and the test sets, $\mathcal{V}_1$ and $\mathcal{V}_2$, have been generated by the stratified sampling coming out from the values belong to the groups $\mathcal{I}_1$ and $\mathcal{I}_2$.

\subsection{Modeling} \label{sec:Modeling}
This study considers carefully selected models, implemented through the MLlib library \cite{Xiangrui2016,Assefi2017}, for the analysis and prediction of Kickstarter campaigns outcomes. The rationale for this choice stems from the lack of algorithms with validated distributed versions in esteemed academic journals. These algorithms, with a strong scientific foundation, offer high transparency and undergo rigorous peer-review processes within the academic community. This allows maintaining a heightened level of confidence in their reliability, ensuring adherence to elevated standards of integrity and quality in their application within the context of such research. Thus, the employed models are:
\begin{itemize}
    \item Decision Tree: a model that partitions the data into smaller subsets based on a set of rules or criteria, see \cite{Pan2009, Irs2022} and its references.
    \item Logistic Regression\footnote{The values of continuous variables have been normalized adopting the standard procedure. \label{normalize_footnote}}: a statistical model that is used to predict a binary outcome based on predictor variables, using a logistic function to convert a linear combination of the input features into a probability value between 0 and 1, see \cite{Yu2012,Lin2014} and its references.
    \item Linear Support Vector Machine (Linear SVM): a supervised learning algorithm that separates data into two classes\footnote{In the multiclasses scenario $\mathcal{P}_1$, the paradigm One-vs-rest is applied.\label{foot_note_multiclass}} by finding the best hyperplane that maximizes the margin between the classes \cite{Graf2004,Lin2014,Sun2023}.
    \item Random Forest: an extension of Bagging that also randomly selects subsets of features used in each data sample \cite{Gen2015}.
    \item Gradient-Boosted Trees (GBTs): builds an additive model in a forward stage-wise fashion; it allows for the optimization of arbitrary differentiable loss functions. In each stage, regression trees are fit on the negative gradient of the loss function, e.g. binary or multiclass\footref{foot_note_multiclass} log loss. Binary classification is a special case where only a single regression tree is induced \cite{Palit2012}.
\end{itemize}
These models are adopted to solve the classification problems, $\mathcal{P}_1$ and $\mathcal{P}_2$, associated to target variables distinguish between $\mathcal{I}_1$ and $\mathcal{I}_2$.

\section{Numerical results}\label{sec:numerical_results}
In this section, the models' performances derived from the conducted experiments and analyses using different metrics: accuracy, precision, recall and F1-score \cite{Has2001} have been reported. The study is focused on both multiclass $\mathcal{P}_1$ and binary $\mathcal{P}_2$ classification problems and the results are presented and commented. Since the target values are unbalanced, all the models, above discussed, have been trained with weighted target data given by equation~\eqref{weight_distribution}.

\subsection{Multiclass classification problem, $\mathcal{P}_1$}
\label{sec:class_multi}
Following the discussion of Section~\ref{sec:Modeling}, the classification problem $\mathcal{P}_1$ considers the multiclass target values in which the outcomes are: Successful, Fail, Canceled and Suspended. The model's results and performances are detailed in the next table:
\begin{table}[ht]
    \centering
    \begin{tabular}{l c c c c}
    \toprule
        \multicolumn{1}{c}{Model} & \multicolumn{4}{c}{Metrics} \\ \cmidrule{2-5}
        & Accuracy & Precision & Recall & F1 \\ \midrule
        Decision Tree & 0.8546   & 0.8774 & 0.8546  & 0.8597\\
        Logistic Regression & 0.7491 & 0.8009 & 0.7491 & 0.7725\\
        Linear SVM & 0.9053 & 0.9068 & 0.9053 & 0.9030\\ 
        Random Forest & 0.7857 & 0.8361 & 0.7857 & 0.7965\\
        GBTs & 0.9423 & 0.9451 & 0.9423 & 0.9378 \\
        \bottomrule        
    \end{tabular}
    \caption{\textit{Multiclass classification problem.} Numerical evidences such as accuracy, precision, recall and F1 for different models like Decision Tree, Logistic Regression, Linear SVM, Random Forest and GBTs applied to classification problem $\mathcal{P}_1$.}
    \label{tab:evidence_multiclass}
\end{table}

Table \ref{tab:evidence_multiclass} displays model metrics for problem $\mathcal{P}_1$, in which GBTs shows the highest performance, followed by Linear SVM and Decision Tree. Remarkable, Logistic Regression, a commonly employed model in crowdfunding literature \cite{zhu2022Proximal}, yields results a bit better than a naive one-half on-half stochastic choice model, and its performance might not justify its widespread usage in such studies. For problem $\mathcal{P}_1$, GBTs model outperforms other models in terms of accuracy, precision, recall, and F1-score within the given configuration. It suggest that such model is well-suited for addressing $\mathcal{P}_1$ and may provide more reliable predictions compared to other models considered. Discussing the results consequences, it's noteworthy to consider the implications for decision-making or practical applications related to $\mathcal{P}_1$. If achieving high metrics is crucial for business objectives, adopting the GBTs model seems to be the most favorable choice. However, if there are constraints or preferences that make such model less suitable, the Linear SVM and Decision Tree models offer viable alternatives.

\subsection{Binary classification problem, $\mathcal{P}_2$}
\label{sec:class_binary}
As introduced in Section~\ref{sec:Modeling}, the classification problem $\mathcal{P}_2$ considers the binary target values in which the outcomes are: Successful and Not Successful. The model's outcomes are shown in the following table:
\begin{table}[ht]
    \centering
    \begin{tabular}{l c c c c}
    \toprule
        \multicolumn{1}{c}{Model} & \multicolumn{4}{c}{Metrics} \\ \cmidrule{2-5}
        & Accuracy & Precision & Recall & F1 \\ \midrule
        Decision Tree & 0.9597 & 0.9634 & 0.9597 & 0.9601\\
        Logistic Regression & 0.9297 & 0.9310 & 0.9296 & 0.9300\\
        Linear SVM & 0.9813 & 0.9818 & 0.9813 & 0.9814\\ 
        Random Forest & 0.8921 & 0.9097 & 0.8921 & 0.8940\\ 
        GBTs & 0.9811 & 0.9818 & 0.9811 & 0.9812\\
        \bottomrule        
    \end{tabular}
    \caption{Numerical evidences such as accuracy, precision, recall and F1 for different models like Decision Tree, Logistic Regression, Linear SVM, Random Forest and GBTs applied to classification problem $\mathcal{P}_2$ \ref{sec:class_binary}.}
    \label{tab:evidence_binary}
\end{table}

Table \ref{tab:evidence_binary} displays model metrics, including accuracy, precision, recall, and F1-score, for $\mathcal{P}_2$ problem. Among the examined models, GBTs and Linear SVM exhibit the highest performances, closely followed by Decision Tree and then Random Forest models. Logistic Regression shows less performance metrics compared to other models, nevertheless such results are much better compared with ones obtained in multiclass problem $\mathcal{P}_1$. GBTs and Linear SVM outperform other models across all metrics, indicating them suitability for $\mathcal{P}_2$'s prediction tasks, unlike Logistic Regression. These findings underscore the importance of considering model choice's implications for decision-making in $\mathcal{P}_2$ applications. Following the argumentation proposed for $\mathcal{P}_1$, employing GBTs or Linear SVM would be advantageous if high metrics are crucial. However, Decision Tree and then Random Forest models offer still viable alternatives. Nevertheless, it is noteworthy that $\mathcal{P}_2$ problem performs better results since binary classification is easier than multiclass one.

\section{Conclusions and future works}\label{sec:conclusions}
In this study, a big data analysis, applied to crowdfunding reward prediction aiming to enhance understanding and prediction accuracy regarding campaign outcomes, has been considered following \cite{Markas2019}. The analyzed big dataset includes campaigns categorized into Success, Failure, Suspended, Canceled, and Live states. By excluding Live projects and formulating two distinct classification problems, namely $\mathcal{P}_1$ and $\mathcal{P}_2$, a class of distributed models has been employed to predict campaign outcomes. Through random stratified sampling, the dataset has been partitioned into training and test sets. Various metrics, including precision, recall, accuracy, and F1-score, were employed to analyze and interpret the results.

Upon evaluating various distributed modeling techniques on these problems, the findings shed light on their efficacy in predicting tasks. Notably, GBTs model emerged as a robust performer across both $\mathcal{P}_1$ and $\mathcal{P}_2$, exhibiting superior metrics compared to other models considered. Conversely, Logistic Regression, often favored in traditional crowdfunding literature, showcased subpar performance, especially in multiclass classification $\mathcal{P}_1$. Nevertheless, in the binary classification problem $\mathcal{P}_2$, Linear SVM and Decision Tree achieve higher results than Logistic Regression under similarly explainable condition. These results, surpassing those reported by \cite{Markas2019}, underscore the importance of model selection in optimizing predictive accuracy, with implications for decision-making in crowdfunding contexts.

The current dataset is limited to a few developed countries, in which the observed indices (GDP and HDI) demonstrate stability, implying a potential restriction in the overall impact on results. However, as the platform will evolve to include projects from developing markets, there is a reasonable expectation of a more substantial influence. This expansion has the potential to enrich the dataset, providing a more comprehensive and globally representative perspective.

Furthermore, the introduced errors at the outset (see Table~\ref{tab:dati_economici}), stemming from the use of approximate data indices, do not significantly influence the classification outcomes and consequently the results.

Looking ahead, several avenues for future research beckon:
\begin{itemize}
    \item Conducting analyses tailored to individual countries can offer valuable insights into localized factors influencing crowdfunding success. By training models on data from specific regions and comparing results, researchers can uncover region-specific trends and dynamics, facilitating more targeted strategies for entrepreneurs and investors alike. Additionally, cross-country model comparisons may unveil transferable insights across different socio-economic contexts. By examining how models trained on one country perform in predicting outcomes in another, researchers can glean insights into universal versus context-specific factors driving campaign success.

    \item Integrating additional data sources (particularly from campaigns started after 01/2018), such as project descriptions, comments, and FAQs, holds promise for enriching predictive models. Although the implementation costs would rise, the potential benefits in terms of enhanced predictive accuracy and deeper understanding of campaign dynamics warrant the further analysis. Moreover, leveraging natural language processing techniques to extract insights from textual data could further refine predictive capabilities.

    \item Developing models to forecast the pledged amount for a campaign, even in its early stages, represents a significant area for future research. By leveraging advanced modeling techniques and real-time data streams, researchers can provide stakeholders with valuable foresight into the potential financial support a campaign is likely to receive, enabling more informed decision-making and strategic planning.

    \item As the crowdfunding platform expands to include projects from developing markets, the dataset’s richness and diversity are expected to increase, potentially offering deeper insights into crowdfunding dynamics on a global scale. Integrating macroeconomic indicators with finer granularity can further enhance predictive models, offering a more nuanced understanding of the interplay between macroeconomic trends and crowdfunding outcomes.

\end{itemize}

\section*{Acknowledgements} 
This research has received funding from the European Union’s NextGenerationUE – Project: Centro Nazionale HPC, Big Data e Quantum Computing, “Spoke 1” (No. CUP E63C22001000006). E. Macca was partially supported by GNCS No. CUP E53C22001930001 Research Project “Metodi numerici per problemi differenziali multiscala: schemi di alto ordine, ottimizzazione, controllo”. E. Macca would like to thank the Italian Ministry of Instruction, University and Research (MIUR) to support this research with funds coming from PRIN Project 2022  (2022KA3JBA, entitled “Advanced numerical methods for time dependent parametric partial differential equations and applications”). E. Macca is member of the INdAM Research group GNCS.

On behalf of all authors, the corresponding author states that there is no conflict of interest.

\bibliography{./biblio}

@article{Mol2014,
	title        = {{The dynamics of crowdfunding: An exploratory study}},
	author       = {Ethan Mollick},
	year         = 2014,
	journal      = {Journal of Business Venturing},
	volume       = 29,
	number       = 1,
	pages        = {1--16},
	issn         = {0883-9026}
}

@inproceedings{Wan2020,
	title        = {{Crowdfunding Dynamics Tracking: A Reinforcement Learning Approach}},
	author       = {Jun Wang and Hefu Zhang and Qi Liu and Zhen Pan and Hanqing Tao},
	year         = 2020,
	booktitle    = {The Thirty-Fourth {AAAI} Conference on Artificial Intelligence, {AAAI} 2020, The Thirty-Second Innovative Applications of Artificial Intelligence Conference, {IAAI} 2020, The Tenth {AAAI} Symposium on Educational Advances in Artificial Intelligence, {EAAI} 2020, New York, NY, USA, February 7-12, 2020},
	publisher    = {{AAAI} Press},
	pages        = {6210--6218},
	bibsource    = {dblp computer science bibliography, https://dblp.org}
}

@article{Song2019,
	title        = {{Mining and investigating the factors influencing crowdfunding success}},
	author       = {Song, Yang and Berger, Ron and  Yosipof,  Abraham and Barnes, Bradley R.},
	year         = 2019,
	journal      = {Technological Forecasting and Social Change Volume 148},
	volume       = 148,
	pages        = {43--58},
	is-sn        = {0040-1625}
}

@article{Wan2022,
	title        = {{The merits of a sentiment analysis of antecedent comments for the prediction of online fundraising outcomes}},
	author       = {Wei Wang a, Lihuan Guo b, Yenchun Jim Wu},
	year         = 2022,
	month        = {01},
	journal      = {Technological Forecasting and Social Change},
}

@article{Zho2022,
	title        = {{Success prediction of crowdfunding campaigns with project network: A machine learning approach}},
	author       = {Zhong, Chao and Xu, Wei and Du,Wei},
	year         = 2022,
	journal      = {Journal of Electronic Commerce Research},
	publisher    = {Journal of Electronic Commerce Research},
	volume       = 23,
	number       = 2,
	pages        = {99--114}
}

@article{blan2022Extracting,
	title        = {{Extracting Image Characteristics to Predict Crowdfunding Success}},
	author       = {Blanchard, S. J. and Noseworthy, T. J. and Pancer, E. and Poole, M.},
	year         = 2022,
	journal      = {arXiv},
	copyright    = {Creative Commons Attribution 4.0 International}
}

@article{zhu2022Proximal,
	title        = {{Proximal language predicts crowdfunding success: Behavioral and experimental evidence}},
	author       = {Xun Zhu},
	year         = 2022,
	journal      = {Computers in Human Behavior},
	volume       = 131,
	pages        = 107213,
	issn         = {0747-5632},
}

@article{Raf2023,
	title        = {{Using machine learning approach towards successful crowdfunding prediction}},
	author       = {Raflesia, Sarifah and Lestarini, Dinda and Kurnia, Rizka and Hardiyanti, Dinna},
	year         = 2023,
	month        = {08},
}

@inproceedings{Koc2016,
	title        = {{The Phenomenon of Project Overfunding on Online Crowdfunding Platforms – Analyzing the Drivers of Overfunding}},
	author       = {Koch, Jascha-Alexander},
	year         = 2016,
	month        = {06},
}

@article{Bolt2020,
  title={Maddison style estimates of the evolution of the world economy. A new 2020 update},
  author={J. Bolt and J.L. {Van Zanden}},
  journal={Maddison-Project Working Paper WP-15, University of Groningen, Groningen, The Netherlands},
  year={2020}
}

@article{Ros2014,
  title={Human development index (HDI)},
  author={Roser, Max},
  journal={Our World in Data},
  year={2014}
}

@article{Zah2016,
author = {M. Zaharia and R.S. Xin and P. Wendell and T. Das and M. Armbrust and A. Dave and X. Meng and J. Rosen and S. Venkataraman and M.J: Franklin and A. Ghodsi and J. Gonzalez and S. Shenker and I. Stoica},
title = {Apache {S}park: {A} {U}nified {E}ngine for {B}ig {D}ata {P}rocessing},
year = {2016},
issue_date = {November 2016},
publisher = {Association for Computing Machinery},
address = {New York, NY, USA},
volume = {59},
number = {11},
journal = {Commun. ACM},
pages = {56–65}
}

@article{King2002,
author = {G. King and L. Zeng},
year = {2002},
month = {09},
pages = {},
title = {Logistic Regression in Rare Events Data},
volume = {9},
journal = {Political Analysis},
doi = {10.1093/oxfordjournals.pan.a004868}
}

@InProceedings{Irs2022,
author="{\.{I}}rsoy, Ozan
and Alpayd{\i}n, Ethem",
editor="Krzyzak Adam and Suen, Ching Y. and Torsello, Andrea and Nobile, Nicola",
title="Distributed Decision Trees",
booktitle="Structural, Syntactic, and Statistical Pattern Recognition",
year="2022",
publisher="Springer International Publishing",
address="Cham",
pages="152--162",
}

@article{Pan2009,
author = {Panda, Biswanath and Herbach, Joshua S. and Basu, Sugato and Bayardo, Roberto J.},
title = {PLANET: Massively Parallel Learning of Tree Ensembles with MapReduce},
year = {2009},
issue_date = {August 2009},
publisher = {VLDB Endowment},
volume = {2},
number = {2},
issn = {2150-8097},
url = {https://doi.org/10.14778/1687553.1687569},
doi = {10.14778/1687553.1687569},
journal = {Proc. VLDB Endow.},
month = {aug},
pages = {1426–1437},
numpages = {12}
}

@article{Yu2012,
  title={Large linear classification when data cannot fit in memory},
  author={Yu, Hsiang-Fu and Hsieh, Cho-Jui and Chang, Kai-Wei and Lin, Chih-Jen},
  journal={ACM Transactions on Knowledge Discovery from Data (TKDD)},
  volume={5},
  number={4},
  pages={1--23},
  year={2012},
  publisher={ACM New York, NY, USA}
}

@inproceedings{Graf2004,
  title={Parallel {S}upport {V}ector {M}achines: {T}he {C}ascade {SVM}},
  author={Hans Peter Graf and Eric Cosatto and L{\'e}on Bottou and Igor Durdanovic and Vladimir Naumovich Vapnik},
  booktitle={Neural Information Processing Systems},
  year={2004}
}

@article{Lin2014,
author = {Lin, Chieh-Yen and Tsai, Cheng-Hao and Lee, Ching-Pei and Lin, Chih-Jen},
year = {2015},
month = {01},
pages = {519-528},
title = {Large-scale logistic regression and linear support vector machines using spark},
journal = {Proceedings - 2014 IEEE International Conference on Big Data, IEEE Big Data 2014},
doi = {10.1109/BigData.2014.7004269}
}

@article{Sun2023,
author = {Sun, Gaoming and Wang, Xiaozhou and Yan, Yibo and Zhang, Riquan},
year = {2023},
month = {08},
title = {Statistical inference and distributed implementation for linear multicategory {SVM}},
volume = {12},
journal = {Stat},
doi = {10.1002/sta4.611}
}

@article{Gen2015,
author = {Genuer, Robin and Poggi, Jean-Michel and Tuleau-Malot, Christine and Villa-Vialaneix, Nathalie},
year = {2015},
month = {11},
title = {Random {F}orests for {B}ig {D}ata},
volume = {9},
journal = {Big Data Research},
doi = {10.1016/j.bdr.2017.07.003}
}

@article{Palit2012,
author = {Palit, Indranil and Reddy, Chandan},
year = {2012},
month = {10},
title = {Scalable and {P}arallel {B}oosting with {M}ap{R}educe},
volume = {24},
journal = {Knowledge and Data Engineering, IEEE Transactions on},
doi = {10.1109/TKDE.2011.208}
}

@article{Belleflamme2014,
    author = {P. Belleflamme and T. Lambert and A. Schwienbacher },
    title = {Crowdfunding: {T}apping the right crowd},
    journal = {Journal of Business Venturing},
    year = 2014
}

@article{Colombo2015,
    author = { M. G. Colombo and C. Franzoni and C. Rossi-Lamastra} ,
    title = { Internal social capital and the attraction of early contributions in crowdfunding},
    journal = {Entrepreneurship Theory and Practice},
    volume = {39(1)},
    pages ={75-100},
    year = 2015 
}

@article{Meer2013,
author = {Meer, J.},
year = {2013},
pages = {},
title = {Effects of the {P}rice of {C}haritable {G}iving: {E}vidence from an {O}nline {C}rowdfunding {P}latform},
volume = {103},
journal = {Journal of Economic Behavior \& Organization}
}

@article{Kraus2016,
  title={Strategies for reward-based crowdfunding campaigns},
  author={Kraus, Sascha and Richter, Chris and Brem, Alexander and Cheng, Cheng-Feng and Chang, Man-Ling},
  journal={Journal of Innovation \& Knowledge},
  volume={1},
  number={1},
  pages={13--23},
  year={2016},
  publisher={Elsevier}
}

@article{Belleflamme2015,
  title={The economics of crowdfunding platforms},
  author={Belleflamme, Paul and Omrani, Nessrine and Peitz, Martin},
  journal={Information Economics and Policy},
  volume={33},
  pages={11--28},
  year={2015},
  publisher={Elsevier}
}

@inproceedings{Etter2013,
  title={Launch hard or go home! {P}redicting the success of {K}ickstarter campaigns},
  author={Etter, Vincent and Grossglauser, Matthias and Thiran, Patrick},
  booktitle={Proceedings of the first ACM conference on Online social networks},
  pages={177--182},
  year={2013}
}

@article{Xing2016,
  title={Strategies and principles of distributed machine learning on big data},
  author={Xing, Eric P and Ho, Qirong and Xie, Pengtao and Wei, Dai},
  journal={Engineering},
  volume={2},
  number={2},
  pages={179--195},
  year={2016},
  publisher={Elsevier}
}

@article{Shneor2020,
  title={Crowdfunding success: a systematic literature review 2010--2017},
  author={Shneor, Rotem and Vik, Amy Ann},
  journal={Baltic Journal of Management},
  volume={15},
  number={2},
  pages={149--182},
  year={2020},
  publisher={Emerald Publishing Limited}
}

@article{Zaharia2010,
author = {Zaharia, Matei and Chowdhury, Mosharaf and Franklin, Michael and Shenker, Scott and Stoica, Ion},
year = {2010},
month = {07},
pages = {10-10},
title = {Spark: Cluster Computing with Working Sets},
volume = {10},
journal = {Proceedings of the 2nd USENIX conference on Hot topics in cloud computing}
}

@inproceedings{Zaharia2012,
author = {Zaharia, Matei and Chowdhury, Mosharaf and Das, Tathagata and Dave, Ankur and Ma, Justin and McCauley, Murphy and Franklin, Michael and Shenker, Scott and Stoica, Ion},
year = {2012},
month = {04},
pages = {2-2},
title = {Resilient distributed datasets: A fault-tolerant abstraction for in-memory cluster computing},
journal = {Proceedings of the 9th USENIX Conference on Networked Systems Design and Implementation}
}

@inproceedings{Assefi2017,
author = {Assefi, Mehdi and Behravesh, Ehsun and Liu, Guangchi and P. Tafti, Ahmad},
year = {2017},
month = {12},
pages = {},
title = {Big Data Machine Learning using Apache Spark MLlib},
doi = {10.1109/BigData.2017.8258338}
}

@article{Xiangrui2016,
  author  = {Xiangrui Meng and Joseph Bradley and Burak Yavuz and Evan Sparks and Shivaram Venkataraman and Davies Liu and Jeremy Freeman and DB Tsai and Manish Amde and Sean Owen and Doris Xin and Reynold Xin and Michael J. Franklin and Reza Zadeh and Matei Zaharia and Ameet Talwalkar},
  title   = {MLlib: Machine Learning in Apache Spark},
  journal = {Journal of Machine Learning Research},
  year    = {2016},
  volume  = {17},
  number  = {34},
  pages   = {1-7},
  url     = {http://jmlr.org/papers/v17/15-237.html}
}

@book{Has2001,
  address = {New York, NY, USA},
  author = {Hastie, Trevor and Tibshirani, Robert and Friedman, Jerome},
  publisher = {Springer New York Inc.},
  series = {Springer Series in Statistics},
  title = {The Elements of Statistical Learning},
  year = 2001
}

@article{Markas2019,
    author = {Markas, Ruhaab and Wang, Yisha} ,
    title = {Dare to {V}enture: {D}ata {S}cience {P}erspective on {C}rowdfunding},
    journal = {SMU Data Science Review},
    volume = {2(1)},
    year = 2019
}

\end{document}